\documentstyle{elsart}

\begin{document}

\begin{frontmatter}

\title{Wide-Field Imaging from Space of
Early-Type Galaxies and Their Globular Clusters}     

\author{Stephen E. Zepf}

\address{Dept. of Physics and Astronomy, Michigan State University, \\
East Lansing, MI 48824, USA}

\begin{abstract}
Wide-field imaging from space will reveal a wealth
of information about the globular cluster systems of any
galaxies in the local universe that are observed by such
a mission. Individual globular clusters around galaxies
in the local universe have compact sizes that are ideal
for the excellent spatial resolution afforded by space-based
imaging, while systems of these globular clusters
have large spatial extent that can only be fully explored
by wide-field imaging. One example of the science 
return from such a study is the determination of the 
major formation epoch(s) of galaxies from the ages 
of their globular clusters determined via their optical to 
near-infrared colors. A second example is determining the 
sites of metal-poor globular cluster formation from
their cosmological bias, which constrains the
formation of structures early in the universe.

\end{abstract}
\end{frontmatter}

\section{Why JDEM Imaging of Globular Cluster Systems}

Globular clusters are invaluable fossil records of the
early formation history of the galaxy in which they are
located. Composed of roughly one million stars that formed 
at the same time with similar composition and that have remained
bound for up to a Hubble time, each globular cluster provides
an observable record of the age, metallicity, and kinematics 
at the time it was formed. Because these quantities are determined 
for individual globulars, studies of globular cluster systems 
can constrain the {\it distribution} of the ages and metallicities
rather than the global average of these key galaxy quantities
revealed by most studies of the integrated light of galaxies.
Globular clusters are also observed to form in major starbursting
episodes in the local universe,
so their properties provide a way to trace the major formation 
episodes in their 
host galaxies. Moreover, because typical globular clusters 
contain roughly one million stars, they are bright and can 
be readily observed as individual objects out to sufficient 
distances ($\sim 15$ Mpc) to obtain a significant sample of 
elliptical galaxies.
This allows the investigation of the formation history of 
massive early-type galaxies that are absent in the Local Group, 
but make up the majority of stellar mass in the local universe
(e.g.\ Hogg et al.\ 2002).
Consequently, studies of globular cluster systems have long
played an important role in constraining how and when the major
formation episodes in galaxies occurred (e.g.\ Ashman \& Zepf 1998,
Harris 2001).

The study of extragalactic globular cluster systems has been
revolutionized by space-based imaging with the Hubble Space
Telescope. This is in large part because the size of extragalactic
globular clusters is very well-matched to diffraction limited 
optical imaging with a 2-m class telescope. Specifically, a 
typical globular cluster half-light radius of several pc
at a distance of 10 Mpc corresponds to $\sim 0.05''$ on the sky. 
However, in one critical aspect, the 
study of extragalactic globular cluster systems is not
well suited to HST imaging. The systems of globular
cluster systems around massive early-type galaxies
extend into the halos of the galaxies,
covering tens of arcminutes on the sky (e.g.\ Rhode \& Zepf 2004
and references therein). Therefore, wide fields of view
are required in order to accurately determine total properties
of globular cluster systems around galaxies.
Moreover, the outer halos of galaxies may hold unique clues
to the assembly history of galaxies, and globular clusters 
are one of the few probes available in these regions.
In this contribution, we highlight two key science questions
that would be addressed by wide-field, space-based imaging
of galaxies in the local universe and their globular cluster
systems.

\section{Constraints on the Formation Epoch(s) of Elliptical Galaxies}

Globular cluster systems are key tools for determining
the formation history of elliptical galaxies 
because the ages and metallicites of their
globular clusters can potentially be determined, thereby 
yielding the major formation epoch(s) of the host galaxies. 
Studies in optical colors
have revealed that globular cluster systems of elliptical
galaxies often have color distributions with two or more
peaks (e.g.\ Kundu \& Whitmore 2001, Larsen et al.\ 2001).
This is one of the clearest signs that elliptical galaxies 
form eposidically, and is consistent with earlier predictions
for elliptical galaxies formed by the mergers of disk galaxies
(Ashman \& Zepf 1992).
While the optical colors of globular cluster systems 
indicate an episodic formation history, 
they do not significantly constrain 
{\it when} these events occur.
This is because optical colors alone can not generally
distinguish between different age and metallicity 
combinations (the age-metallicity degeneracy).
In some specific cases with strongly bimodal color
distributions, the cluster systems must be mostly old and 
have a bimodal metallicity distribution to account for
the red and blue populations (e.g.\ Zepf \& Ashman 1993).
However, a general understanding of the age and metallicity 
distribution of globular cluster systems 
requires the breaking of the
age-metallicity degeneracy.

The addition of near-infrared photometry to optical data
provides a way to break the age-metallicity degeneracy
(e.g.\ Puzia et al.\ 2002). The basis for this technique
is that the flux of a simple stellar population in the 
near-infrared is primarily sensitive to metallicity, while
optical fluxes have a greater sensitivity to age.
This approach has resulted in the first identification of
a substantial population of intermediate-age globular clusters 
in an ordinary elliptical galaxy 
(Puzia et al.\ 2002), 
and studies are now being carried out of larger samples of 
galaxies (e.g.\ Hempel et al.\ 2003). However, these studies 
are limited to the inner regions of galaxies because of
the modest size of current near-infrared arrays and
also, more fundamentally, because of the low surface density 
of globular clusters in the outer regions of galaxies. 
Specifically, even with careful image classification and 
multicolor selection, contamination of globular cluster 
samples in wide-field ground-based imaging is a concern 
(e.g.\ Rhode \& Zepf 2001). As demonstrated by HST, 
imaging at 
$0.1''$ resolution effectively eliminates 
most of this concern by separating background galaxies,
foreground stars, and globular clusters 
(e.g.\ Kundu et al.\ 1999).

Any JDEM observations of early-type galaxies in the local universe 
would provide optical and near-infrared luminosities of their
globular clusters over a wide field with little contamination.
This would then allow the use of the optical to near-infrared
technique to determine the ages and metallicities of the major 
formation epoch(s) of these galaxies from the inner regions 
out into their halos. This would place important constraints
on the formation history of elliptical galaxies, particularly
in their outer regions for which there are few other constraints.

\section{Cosmological Bias of Metal-Poor Globular Cluster Populations}
The low metallicity and the ``halo''-like extended spatial distribution
of metal-poor globular cluster systems suggests that they 
generally formed at high redshift. 
These objects may therefore be the
best available fossil records of structure formation at early epochs. 
One way to address the question of the formation sites of metal-poor 
globular clusters is to determine their ``cosmological bias.'' 
Specifically, galactic halos that
are more massive and in denser environments will tend to have
more of their mass collapsed by a given redshift than halos
of lower mass and in poorer environments. Thus, a given 
formation redshift for metal-poor globular clusters translates 
directly into a specific prediction for the `biasing'' of 
metal-poor clusters towards higher mass halos, with the adoption
of a constant formation efficiency of globular clusters per
collapsed halo mass.
This can be further tested by comparison with the spatial
distribution of the metal poor clusters, which
is also dependent on formation epoch and halo mass
(see Rhode \& Zepf 2004 and Santos 2003).

To determine the cosmological bias of the metal-poor globular
clusters, their total number and spatial distribution around
galaxies of different masses and environments are required.
Current work along these lines has primarily utilized the
ground-based CCD Mosaic cameras covering fields of roughly 
$30' \times 30'$ (e.g.\ Rhode \& Zepf 2004). These data are 
a significant improvement over those available earlier, and 
the field size covers most or all of the
metal-poor globular cluster systems of a broad range of galaxies
in the local universe. However, as discussed in the previous
section, space-based imaging is invaluable for identifying
globular clusters over these large fields. Therefore, JDEM imaging
of galaxies in the local universe would provide high 
quality total numbers and spatial profiles of the metal-poor component
of globular cluster systems for the determination of the biasing of 
this population. This would then constrain when and where these 
structures formed early in the universe.

I would like to thank my many collaborators on projects
related to those described above, and acknowledge support from
NASA Long-Term Space Astrophysics grant NAG 5-11319.

\end{document}